\documentclass[11pt,notitlepage,nofootinbib]{revtex4-1}

\usepackage{epsf}
\usepackage{latexsym}
\usepackage{amssymb}
\usepackage{amsmath}
\usepackage{bm}
\usepackage[dvips]{graphicx}
\usepackage{array}
\usepackage{xcolor}

\date{\today}
\def\d3{^{(3)}\nabla}

\begin{document}

\title{Cross correlations of the CMB Doppler mode and the 21 cm brightness temperature in the presence of a primordial magnetic field}

\author{Kerstin E. Kunze}

\email{kkunze@usal.es}

\affiliation{Departamento de F\'\i sica Fundamental, Universidad de Salamanca,
 Plaza de la Merced s/n, 37008 Salamanca, Spain}

\begin{abstract}
The cross correlation between the CMB Doppler mode and the 21 cm line brightness temperature is calculated in the presence of a stochastic primordial magnetic field. 
Potential detectability is estimated for Planck 2018 bestfit parameters in combination with configuration and survey design parameters of 
21 cm line radio telescopes such as LOFAR and  the future SKAO.
Homogeneous as well as inhomogeneous reionization has been considered. In particular the latter in combination with SKA1-mid shows promising signal-over-noise ratios. 
\end{abstract}

\maketitle

\section{Introduction}
\label{s0}
\setcounter{equation}{0}
There is ample evidence for magnetic fields being present in galaxies and clusters of galaxies and even in filaments on scales beyond clusters. All of these observations associate magnetic fields 
with collapsed structures. Since about  the last decade observational data of a number of blazars have been interpreted as evidence of the  presence of magnetic fields in voids. This could indicate the existence of cosmological magnetic fields not associated with any virialized structure but rather pervading the observable universe as a whole.
These latter observations together with for example the  question of the origin of galactic magnetic fields 
motivate the consideration  of a primordial magnetic field (for reviews, e.g., \cite{Durrer:2013pga,rev4}).

If these large scale magnetic fields are generated in the very early universe before decoupling cosmological observables such as, e.g., the temperature anisotropies and polarization of the cosmic microwave background (CMB) are affected.
There are several ways how cosmological magnetic fields enter into the dynamics of cosmological perturbations that eventually leave their imprint on the angular power spectra of the CMB (cf. e.g., \cite{Giovannini:2004aw, Giovannini:2008aa,pfp,kb,sl,kk11,sl12,kk12}).
Directly magnetic fields contribute to the total energy and anisotropic stress perturbations as well as to the evolution of the baryon velocity perturbations. Indirectly they have an important effect because of their dissipation due to the interaction with the cosmic plasma. This leads to additional energy injection and heating of matter \cite{sesu,kuko15}.

Motivated by the observed isotropy of the universe on very large scales  primordial magnetic fields  are assumed to be gaussian random fields.
Various observations, such as the CMB, limit their present day amplitude to be of order of nG or below (cf. e.g. \cite{Planck2015cmf}).
Current and upcoming observations of the 21 cm line of neutral hydrogen provide new opportunities of constraining the parameters of a putative primordial magnetic field.

Over the last three decades observations of the CMB have opened a window to the physics of the very early universe. The angular power spectrum of the temperature anisotropies, the E-mode of polarization and their cross correlation have been measured with unprecedented precision. The search for a putative primordial B-mode of the linear polarization is still going on. Since the first observations of quasars and the Gunn-Petersen trough it is known that  the universe at least since a redshift of $z=6$ is reionized. Observations of the CMB temperature anisotropies and polarization 
such as the Planck 2018 data indicate a best fit value for the $\Lambda$CDM base model of  $z=7.68$ \cite{Planck-2018}.
This assumes  nearly instanteneous  reionization.
Observations of the 21 cm line signal resulting from the hyperfine transition of neutral hydrogen measured against the CMB radiation is very sensitive to the epoch of reionization. Thus  current and upcoming experiments dedicated to the observation of the cosmic 21 cm line signal  will provide  an insight into the details of the reionization epoch. Moreover, since the neutral hydrogen distribution traces the total matter distribution in the universe also details of the latter will  be open to observations. In addition taking into account the redshifting of the 21 cm line signal observations at different frequencies
correspond to different redshifts and thus different stages in the evolution of the universe (for reviews, e.g., \cite{FurlanettoPRep, LoebFurl1stGal}).
Since primordial magnetic fields have a direct effect on  the linear matter power spectrum on small scales the cosmic 21 cm line signal provides an interesting probe into their existence. In \cite{kk2019} the resulting 21 cm line signal has been obtained from simulations with the {\tt Simfast21} \footnote{https://github.com/mariogrs/Simfast21} code  \cite{simfast21-1,simfast21-2}  using this modified linear matter power spectrum as initial condition.

One of the effects of reionization is the rescattering of photons leading to a secondary Doppler contribution in the CMB anisotropies which is determined by the matter power spectrum. The cross correlation with the 21 cm signal has been first studied in \cite{Alvarez:2005sa}  with the objective to  constrain the parameters of the reionization model and subsequently in \cite{Adshead:2007ij}.

The aim here is to study this cross correlation in the presence of a primordial magnetic field  and the prospects  to constrain the magnetic field parameters using the characteristics of  radio telescope arrays  such as  
the upcoming Square Kilometre Array Observatory (SKAO) with the updated specifications \cite{SKAtelecon} as well as the  already operational Low Frequency Array (LOFAR) \cite{2013LOFAR}.
In the numerical solutions the best fit parameters of the base model derived from Planck 2018 data only are used \cite{Planck-2018}, in particular, 
$\Omega_bh^2$ = 0.022383, $\Omega_m h^2$ =0.14314, $H_0 = 67.32 $ km s$^{-1}$Mpc$^{-1}$ and for the adiabatic mode
$A_s = 2.101\times 10^{-9}$ and  $n_s = 0.96605$. The angular power spectra of the CMB temperature anisotropies, the linear matter power spectra as well as other required functions of the cosmological background  in the numerical solutions  are calculated for the adiabatic mode and the compensated magnetic mode with a modified version of the CLASS code \cite{kk21} (for references of the original CLASS code cf  \cite{class1,class2,class3,class4,class5}).

\section{The signals}
\label{s1}
\setcounter{equation}{0}
Observations of the CMB indicate that the largest contribution to the  total density perturbation is due to the adiabatic, primordial curvature mode (e.g.,\cite{Planck-2018}).
 For the magnetic field contribution initial conditions corresponding to the compensated magnetic mode are chosen   for the CMB Doppler mode as well as the linear matter power spectrum. For this type of initial condition the neutrino anisotropic stress approaches after neutrino decoupling a solution compensating the contribution of the anisotropic stress of the magnetic field \cite{sl, kk11,kk12}. Having a behaviour similar to an isocurvature mode its contribution has to be subleading in comparison to the adiabatic mode.

The magnetic field ${\bf B}$ is assumed to be a non helical, gaussian random field determined by its two point function in $k$-space,
\begin{eqnarray}
\langle B_i^*(\vec{k})B_j(\vec{q})\rangle=(2\pi)^3\delta({\vec{k}-\vec{q}})P_B(k)\left(\delta_{ij}-\frac{k_ik_j}{k^2}\right),
\end{eqnarray}
where the power spectrum, $P_B(k)$ is given by \cite{kk11}
\begin{eqnarray}
P_B(k,k_m,k_L)=A_B\left(\frac{k}{k_L}\right)^{n_B}{\cal W}(k,k_m)
\end{eqnarray}
with $A_B$ its amplitude, $k_L$  a pivot wave number chosen to be 1 Mpc$^{-1}$ and  ${\cal W}(k,k_m)=\pi^{-3/2}k_m^{-3}e^{-(k/k_m)^2}$
 a gaussian window function. $k_m$ corresponds to the largest scale
damped due to radiative viscosity before decoupling \cite{sb,jko}. $k_m$ has its largest value at recombination 
\begin{eqnarray}
k_m=301.45\left(\frac{B}{\rm nG}\right)^{-1}{\rm Mpc}^{-1}
\end{eqnarray}
for the bestfit parameters of Planck 2018 data only \cite{kk21,Planck-2018}.

\subsection{The 21 cm line signal: Auto correlation}

The cosmic 21 cm line signal corresponds to the change in the brightness temperature of the CMB
which in general  depends on direction ${\bf n}$ and redshift $z$.
An overline indicates the corresponding homogeneous background variable.
Expanding the differential brightness temperature of the 21 cm signal upto first order, 
$T_{21}({\bf n},z)=\overline{T}_{21}(z)+\delta T_{21}({\bf n},z)$,
the observed 
differential brightness perturbation corresponding to the cosmic 21 cm line signal  originating at a redshift $z$ is given by \cite{Zaldarriaga:2003du, Alvarez:2005sa,Adshead:2007ij} 
\begin{eqnarray}
T_{21}({\bf n},z)=T_0(z)\int_{0}^{\eta_0}d\eta' W\left[\eta(z)-\eta'\right]\psi_{21}({\bf n},\eta'),
\end{eqnarray}
where $\eta$ is conformal time and $W(x)$ denotes the spectral response function of the instrument which for simplicity is chosen to be a $\delta$ function, $W(x)=\delta(x)$.  The signal generated by regions of neutral hydrogen at redshift $z$ will be observed today at $\nu_{21 cm}=1420.4057\;{\rm MHz}/(1+z)$.
Moreover \cite{Adshead:2007ij} ,
\begin{eqnarray}
T_0(z)&=&23 {\rm mK}\left(\frac{\Omega_bh^2}{0.02}\right)\left[\left(\frac{0.15}{\Omega_mh^2}\right)
\left(\frac{1+z}{10}\right)\right]^{\frac{1}{2}},\\
\psi_{21}({\bf n},\eta),&=&x_{H}({\bf n},\eta)\left(1+\delta_b({\bf n},\eta)-\frac{1}{a(\eta)H(\eta)}\frac{\partial v_r}{\partial r}\right)
\left(1-\frac{T_{CMB}}{T_s({\bf n},\eta)}\right),
\label{psi21}
\end{eqnarray}
where $\delta_b$ is the baryon energy density contrast, $x_H$ refers to the fraction of neutral hydrogen and as in \cite{Adshead:2007ij} 
the effect of redshift space distortion upto linear order is included.
This is due to the fact that the peculiar velocity of distant, massive objects  leads to a distortion of their position in redshift space \cite{Kaiser:1987qv, Shaw:2008aa}.

From around the beginning of reionization it  can be assumed that the spin temperature $T_s$ of neutral hydrogen is much higher than the one of the CMB.
Using equation (\ref{psi21}) together 
with $\psi({\bf n},z)=\overline{\psi}_{21}(z)+\delta\psi_{21}({\bf n},z)$ implies
$\overline{T}_{21}(z)=T_0(z)x_H(z)$ and $\delta T_{21}({\bf n},z)=T_0(z)\delta\psi_{21}({\bf n},\eta)$.
Taking into account the possibility of inhomogeneous reionization then
\begin{eqnarray}
\delta\psi_{21}({\bf n},\eta)= D(\eta)\left[\bar{x}_H(\eta)\delta_b({\bf n,\eta_0})
-\bar{x}_e\delta_x({\bf n,\eta_0})\right]-\frac{\overline{x}_{H}(\eta)}{a(\eta)H(\eta)}\frac{\partial v_r}{\partial r}
\end{eqnarray}
where $D(\eta)$ is the growth factor of the baryon density perturbation $\delta_b$ during matter domination.
Moreover, it is assumed that the time evolution of the ionization fraction perturbation $\delta_x=(x_e-\bar{x}_e)/\bar{x}_e$ follows that of  the linear matter perturbation.
The total baryon density perturbation is given by $\delta_{b,tot}=\delta_{b,ad}+\delta_{b,B}$ including both the adiabatic primordial curvature mode and the magnetic mode. 
During matter domination the evolution of the energy density perturbation of the magnetic mode follows that of the adiabatic mode so that $\delta_{b,tot}({\bf k},\eta)=D(\eta)\delta_{b,tot}({\bf k},\eta_0)$ which follows the evolution of the 
total matter density perturbation $\delta_{m,tot}$ (e.g. \cite{Kunze:2013iwa}).  Finally, the baryon velocity perturbation is determined by $v=-\dot{\delta}_b/k$ \cite{kodamaSasaki}.

Following  \cite{Zaldarriaga:2003du, Alvarez:2005sa,Adshead:2007ij} 
 expanding $\delta\psi_{21}({\bf n},\eta)$ in terms of spherical harmonics (e.g. \cite{hw}) yields  the corresponding expansion coefficients for the 21 cm line signal brightness perturbation as,
\begin{eqnarray}
a^{(21)}_{\ell m}&=&4\pi(-i)^{\ell}T_0(z)\sum_{\bf k}
\left[D(\eta)\left[\bar{x}_H(\eta)\delta_b({\bf k})-\bar{x}_e(\eta)\delta_x({\bf k})\right]
j_{\ell}(k(\eta_0-\eta(z)))
\right.
\nonumber\\
&&+\left.
\frac{\bar{x}_H(\eta) k}{a(\eta)H(\eta)}j_{\ell}''(k(\eta_0-\eta(z)))v({\bf k},\eta)
\right]
Y_{\ell m}^*({\bf\hat{k}}),
\label{a21}
\end{eqnarray}
where 
\begin{eqnarray}
j_{\ell}''(w)=\frac{\ell(\ell-1)}{(2\ell+1)(2\ell-1)}j_{\ell-2}(w)-\frac{2\ell^2+2\ell-1}{(2\ell-1)(2\ell+3)}j_{\ell}(w)+
\frac{(\ell+1)(\ell+2)}{(2\ell+1)(2\ell+3)}j_{\ell+2}(w).
\end{eqnarray}

In general the correlation functions of two modes $X$ and $Y$, respectively, are defined in terms of angular power spectra $C_{\ell}^{(XY)}$ such that 
\begin{eqnarray}
\langle a^{(X)}_{\ell' m'}a^{(Y)*}_{\ell m}\rangle= C^{(XY)}_{\ell}\delta_{\ell \ell'}\delta_{m m'}.
\end{eqnarray}

The autocorrelation function of the 21 cm line signal from redshift $z$ is determined by the angular power spectrum 
\begin{eqnarray}
C_{\ell}^{(21-21)}(z)&=&\frac{2}{\pi}T_0(z)^2D(\eta)^2
\nonumber\\
&\times&
\int dk k^2\left[
\left[\bar{x}_H(z)^2P_{m,tot}(k)
-2\bar{x}_H(z)\bar{x}_e(z)P_{\delta_m \delta x}(k)
+\bar{x}_e(z)^2P_{\delta x\delta x}(k)\right]j_{\ell}(w)j_{\ell}(w)
\right.
\nonumber\\
&-& \left.
2\bar{x}_H(z)\left[\bar{x}_H(z)P_{m,tot}(k)
-\bar{x}_e(z)P_{\delta_m \delta x}(k)\right]j_{\ell}(w)j_{\ell}''(w)
\right.
\nonumber\\
&+&\left.
\bar{x}_H(z)^2
P_{m,tot}(k)j_{\ell}''(w)j_{\ell}''(w)
\right],
\end{eqnarray}
where $w=k(\eta_0-\eta(z))$.
Moreover, $P_{\delta_m\delta x}(k)$ denotes the cross correlation 
power spectrum of the ionized fraction and matter density perturbations. The autocorrelation power spectrum of the 
matter density perturbation is given by $P_{m}(k)$ and of the
ionization fraction perturbation  by $P_{\delta x \delta x }(k)$.  

In the approximation for 
large $\ell$ it is found that
\begin{eqnarray}
&C_{\ell}^{(21-21)}(z)&=\frac{2T_0(z)^2D(\eta)^2}{[\eta_0-\eta(z)]^3}
\left[
\frac{x_{\ell,max}^{(1)}}{4}
\left[\bar{x}_H(z)^2P_{m,tot}(k_{\ell}^{(1)})
-2\bar{x}_H(z)\bar{x}_e(z)P_{\delta_m \delta x}(k_{\ell}^{(1)})
\right.\right.
\nonumber\\
&&\left.\left.\hspace{4.2cm}
+\bar{x}_e(z)^2P_{\delta x\delta x}(k_{\ell}^{(1)})\right]
\right.
\nonumber\\
&-&
\left.
\frac{[x_{\ell,max}^{(2)}]^2(1-2\ell(\ell+1))}{(3+2\ell)(4\ell^2-1)}
\bar{x}_H(z)\left[\bar{x}_H(z)P_{m,tot}(k_{\ell}^{(2)})
-\bar{x}_e(z)P_{\delta_m \delta x}(k_{\ell}^{(2)})\right]
\right.
\nonumber\\
&+&
\left.
\frac{3[x_{\ell,max}^{(3)}]^2[3+2\ell(\ell+1)(\ell^2+\ell-4)]}
{2(4\ell^2-9)(4\ell-1)(5+2\ell)}\bar{x}_H(z)^2
P_{m,tot}(k_{\ell}^{(3)})
\right]
\label{C21-21}
\end{eqnarray}
where  $k_{\ell}^{(i)}=x^{(i)}_{\ell,max}/(\eta_0-\eta(z))$ for $i=1,2,3$
 and
$x^{(1)}_{\ell,max}=\ell+0.5$, 
$x^{(2)}_{\ell,max}=\ell+0.5+0.7(\ell+0.5)^{\frac{1}{3}}$
and 
$x^{(3)}_{\ell,max}=\ell+0.5+1.61723(\ell+0.5)^{\frac{1}{3}}$
obtained from numerical fits to the maximum of the relevant products 
$j_{\ell_1}j_{\ell_2}$.

\begin{figure}[h!]
\centerline{\epsfxsize=3.2in\epsfbox{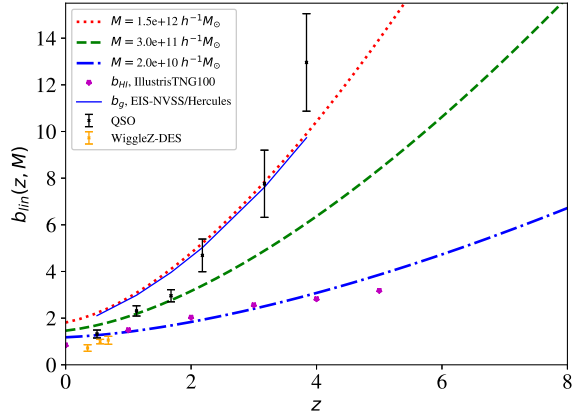}}
\caption{Numerical fitting function for the linear bias function b(z) (\ref{bias}) \cite{Basilakos, Papageorgiou} for different values of the halo mass $M_h$ 
for the best fit parameters of Planck 2018 data  \cite{Planck-2018}.
Data points  for the  HI bias from the Illustris/TNG 100 numerical simulations \cite{illustris} are included. 
A numerical fitting function for the galactic bias derived from observations of radio galaxies of the combined EIS-NVSS and Hercules surveys is also shown 
\cite{EIS-NVSS-Hercules}
as well as galactic bias derived from observations of quasars drawn from the SDSS DR5  quasar catalog \cite{qso} and galaxies from the WiggleZ Dark Energy Survey \cite{wiggleZ}.}
\label{fig0}
\end{figure}

For homogeneous reionization $P_{\delta_m\delta x}(k)$ and  $P_{\delta x \delta x }(k)$ are identically zero. 
However, density perturbations of the background cause reionization  to be inhomogeneous.  
In \cite{Adshead:2007ij} an effective description of reionization is used in which high density regions, described by bubbles,  are ionized first. This is based on 
numerical simulations of reionization being driven by photon emission of star forming galaxies.
In this case the ionization fraction perturbation $\delta x$ is determined by  \cite{Adshead:2007ij}
\begin{eqnarray}
 \delta x(\mathbf{k},z)=b(z)\delta_m(\mathbf{k},z)(1-f_{sat})
\label{deltax}
\end{eqnarray}
where $b(z)$ is the average galaxy bias function at redshift $z$ and $f_{sat}$ is the fraction of saturated bubbles, that is, those that  have reached their maximal radius beyond which recombination within the enclosed region sets in.
This is modelled by  \cite{Adshead:2007ij,2005MNRAS.363.1031F}
\begin{eqnarray}
f_{sat}(\bar{x}_e)=\exp\left(\frac{|\bar{x}_e-1|}{\Delta \bar{x}_e}\right)
\end{eqnarray}
with  $\Delta \bar{x}_e$ a model parameter encapsulating the contribution of recombining regions within bubbles extended beyond their maximal radius. In the numerical solutions presented here the latter is set to  $\Delta \bar{x}_e=0.2$.
Moreover, $\delta_m(\mathbf{k},z)$ is the total matter density perturbation.
Thus the cross correlation function between the ionization fraction perturbation and the density mode at equal redshift is given by,
\begin{eqnarray}
P_{\delta_m\delta x}(k,z)=\langle \delta_m(\mathbf{k},z), \delta x(\mathbf{k}',z)\rangle=b(z)(1-f_{sat})P_m(k,z)\delta_{\mathbf{k},\mathbf{k}'}
\label{pdmdx}
\end{eqnarray}
and analogously the ionization fraction perturbation auto correlation function is given by
\begin{eqnarray}
P_{\delta x\delta x}(k,z)=\langle \delta x(\mathbf{k},z) \delta x(\mathbf{k}',z)\rangle=b^2(z)(1-f_{sat})^2P_m(k,z)\delta_{\mathbf{k},\mathbf{k}'}.
\label{pdxdx}
\end{eqnarray}
The linear bias function is assumed to be given by the numerical fitting function \cite{Basilakos, Papageorgiou}
\begin{eqnarray}
b(z)=C_1E(z)+C_2E(z)I(z)+1
\label{bias}
\end{eqnarray}
where $E(z)=\left(\Omega_m(1+z)^3+\Omega_{\Lambda}\right)^{\frac{1}{2}}$ and  $I(z)=\int_z^{\infty}\frac{(1+x)^3}{E^3(x)}dx$. The latter  can be expressed in terms of the hypergeometric function $_2F_1(a,b;c;z)$ as \cite{AS,Basilakos}
\begin{eqnarray}
I(z)=\frac{2}{\Omega_m^{\frac{3}{2}}}(1+z)^{-\frac{1}{2}}\, _2F_1\left(\frac{3}{2},\frac{1}{6};\frac{7}{6};-\frac{\Omega_{\Lambda}}{\Omega_m}\frac{1}{(1+z)^3}
\right).
\end{eqnarray}
Finally, $C_1$ and $C_2$ are two model parameters depending on the halo mass $M_h$  as
\begin{eqnarray}
C_i(M_h)=\alpha_i\left(\frac{M_h}{10^{13}h^{-1}M_{\odot}}\right)^{\beta_i}\hspace{2cm} i=1,2.
\end{eqnarray}
The parameters $\alpha_i$ and $\beta_i$ are given by \cite{Papageorgiou}
\begin{eqnarray}
\alpha_1&=&4.53\left(\frac{0.81}{\sigma_8}\right)^{\kappa_1}\exp\left[\kappa_2(\Omega_m-0.273)\right], \hspace{0.5cm}
(\kappa_1,\kappa_2)=\left\{
\begin{array}{ll}
(12.15,0.30) & \Omega_m\leq 0.273\\
(8.70,0.37)& \Omega_m>0.273
\end{array}
\right.\\
\alpha_2 &=&-0.41\left(\frac{0.273}{\Omega_m}\right)^n\hspace{1cm} n\simeq 1.4\\
\beta_1&\simeq&\beta_2\simeq 0.35
\end{eqnarray}
In figure \ref{fig0} the bias numerical fitting function  $b(z)$ (\ref{bias}) for different values of the halo mass $M_h$ is shown for the best fit parameters of Planck 2018 data  \cite{Planck-2018} together with data points and numerical fitting functions from numerical simulations \cite{illustris} as well as observations \cite{EIS-NVSS-Hercules,qso,wiggleZ}. In the numerical solutions $M_h=1.5\times 10^{12} h^{-1}M_{\odot}$ is chosen.

\subsection{The CMB Doppler-21 cm line signal cross correlation}

Moving ionized matter and in particular electrons cause temperature fluctuations. The resulting primary CMB anisotropies have  a maximum at 
around multipoles  $\ell\simeq{\cal O}(100)$. However, the corresponding Doppler term in the final line-of-sight integral of the brightness perturbation is strongly damped for large multipoles due to rapid oscillations in the integrand when considering a homogeneous electron distribution \cite{hw}.
Secondary CMB anisotropies  are generated for these large multipoles  at around the epoch of reionization when taking into account the perturbations in the baryon energy density inducing electron number density fluctuations. This is the  Ostriker-Vishniac effect \cite{vishniac,ov,hu,hw1}. In the case of a magnetized primordial plasma there is an additional contribution to these secondary temperature anisotropies since in this case the velocity perturbation not only receives a contribution from the scalar mode but from the vector mode as well \cite{Kunze:2013iwa}.

Following \cite{Alvarez:2005sa,Adshead:2007ij} electron number density fluctuations in the Doppler mode will not be taken into account for the CMB Doppler-21 cm line cross correlation function. Hence the 
brightness perturbation due to the Doppler mode is given by \cite{hw} 
\begin{eqnarray}
\delta T^{(D)}({\bf n}, {\bf x},\eta)=\sum_{\ell,m}a^{(D)}_{\ell m}Y_{\ell m}({\bf n})
\label{TD}
\end{eqnarray}
with  expansion coefficients  
 \begin{eqnarray}
 a^{(D)}_{\ell m}=4\pi (-i)^{\ell} T_{CMB}
 \int\frac{d^3{\bf k}}{(2\pi)^3}\frac{\delta_{b}({\bf k},\eta_0)}{k^2}Y^*_{\ell m}({\bf k})\int_0^{\infty}d\eta g(\eta)\dot{D}(\eta)\frac{\partial j_{\ell}\left(k\left(\eta_0-\eta
 \right)\right)}{\partial\eta}
  \label{aD}
 \end{eqnarray}
 where  $T_{CMB}$ is the temperature of the CMB today, $g(\eta)$ is the visibility function
 and a dot indicates derivative w.r.t. conformal time $\eta$.
For $\delta_{b}({\bf k},\eta_0)=\delta_{b,tot}({\bf k},\eta_0)$ contributions from both the primordial curvature, adiabatic mode and the  magnetic mode are included.

The cross correlation between the CMB Doppler contribution and the 21 cm line
signal is calculated in the approximation for large $\ell$ \cite{Alvarez:2005sa,Adshead:2007ij}.
Following  \cite{hu,hw1,Kunze:2013iwa} integrals over products of comparatively slowly varying functions ${\cal F}(w)$ and spherical Bessel functions are approximated by using
\begin{eqnarray}
\int_0^{\infty}dw {\cal F}(w)j_{\ell}(w)\simeq {\cal F}(w_{\ell})\int_0^{\infty}dw j_{\ell}(w)=
\frac{\sqrt{\pi}}{2}\frac{\Gamma\left(\frac{\ell +1}{2}\right)}{\Gamma\left(\frac{\ell+2}{2}\right)} {\cal F}(w_{\ell})
\end{eqnarray}
where $w_{\ell}=\ell +\frac{1}{2}$ is the position of the first maximum of $j_{\ell}(w)$ \cite{AS,GradRhyz}.
This yields the angular power spectrum of the cross correlation two point function,
\begin{eqnarray}
C_{\ell}^{(D-21)}(z)=-\frac{2}{\pi}T_0(z)T_{CMB}D(z){\cal N}_{\beta}
\left[\bar{x}_H(z)\sum_{i=1}^{3}I_i+\bar{x}_e(z)J\right],
\end{eqnarray}
defining 
\begin{eqnarray}
{\cal N}_{\beta}&=&\frac{\sqrt{\pi}}{2}\frac{\Gamma\left(\frac{\ell +1}{2}\right)}
{\Gamma\left(\frac{\ell+2}{2}\right)}\nonumber\\
I_1&=&-\frac{\sqrt{\pi}}{2}\frac{\ell(\ell-1)}{(2\ell-1)(2\ell+1)(\ell-\frac{3}{2})}
\frac{\Gamma\left(\frac{\ell-1}{2}\right)}{\Gamma\left(\frac{\ell}{2}\right)}
\beta_0(\eta_{\ell}^{(1)})P_{m,tot}(k_{\ell}^{(1)})\nonumber\\
I_2&=&\frac{\sqrt{\pi}}{2}
\frac{2(3\ell^2+3\ell-2)}{(2\ell-1)(2\ell+3)(\ell+\frac{1}{2})}
\frac{\Gamma\left(\frac{\ell+1}{2}\right)}{\Gamma\left(\frac{\ell+2}{2}\right)}
\beta_0(\eta_{\ell}^{(2)})P_{m,tot}(k_{\ell}^{(2)})\nonumber\\
I_3&=&-\frac{\sqrt{\pi}}{2}
\frac{(\ell+1)(\ell+2)}{(2\ell+1)(2\ell+3)(2\ell+\frac{5}{2})}
\frac{\Gamma\left(\frac{\ell+3}{2}\right)}{\Gamma\left(\frac{\ell+4}{2}\right)}
\beta_0(\eta_{\ell}^{(3)})P_{m,tot}(k_{\ell}^{(3)})\nonumber\\
J&=&\frac{\sqrt{\pi}}{2}\frac{\Gamma\left(\frac{\ell+1}{2}\right)}{\Gamma\left(\frac{\ell+2}{2}\right)}
\frac{\beta_0(\eta_{\ell}^{(2)})}{\ell+\frac{1}{2}}
P_{\delta_m\delta_x}(k_{\ell}^{(2)}),
\label{CD-21}
\end{eqnarray}
with the approximations $\eta_{\ell}^{(1)}=\frac{(\ell+\frac{1}{2})\eta(z)-2\eta_0}
{\ell-\frac{3}{2}}$, $k_{\ell}^{(1)}=\frac{\ell-\frac{3}{2}}{\eta_0-\eta(z)}$ and $\eta_{\ell}^{(2)}=\eta(z)$, $k_{\ell}^{(2)}=\frac{\ell+\frac{1}{2}}{\eta_0-\eta(z)}$ and finally
 $\eta_{\ell}^{(3)}=\frac{(\ell+\frac{1}{2})\eta(z)+2\eta_0}
 {\ell+\frac{5}{2}}$, $k_{\ell}^{(3)}=\frac{\ell+\frac{5}{2}}{\eta_0-\eta(z)}$.
Moreover, $\beta_0(\eta)=\frac{d}{d\eta}\left[\dot{D}(\eta)g(\eta)\right]$. 

\begin{figure}[h!]
\centerline{\epsfxsize=2.8in\epsfbox{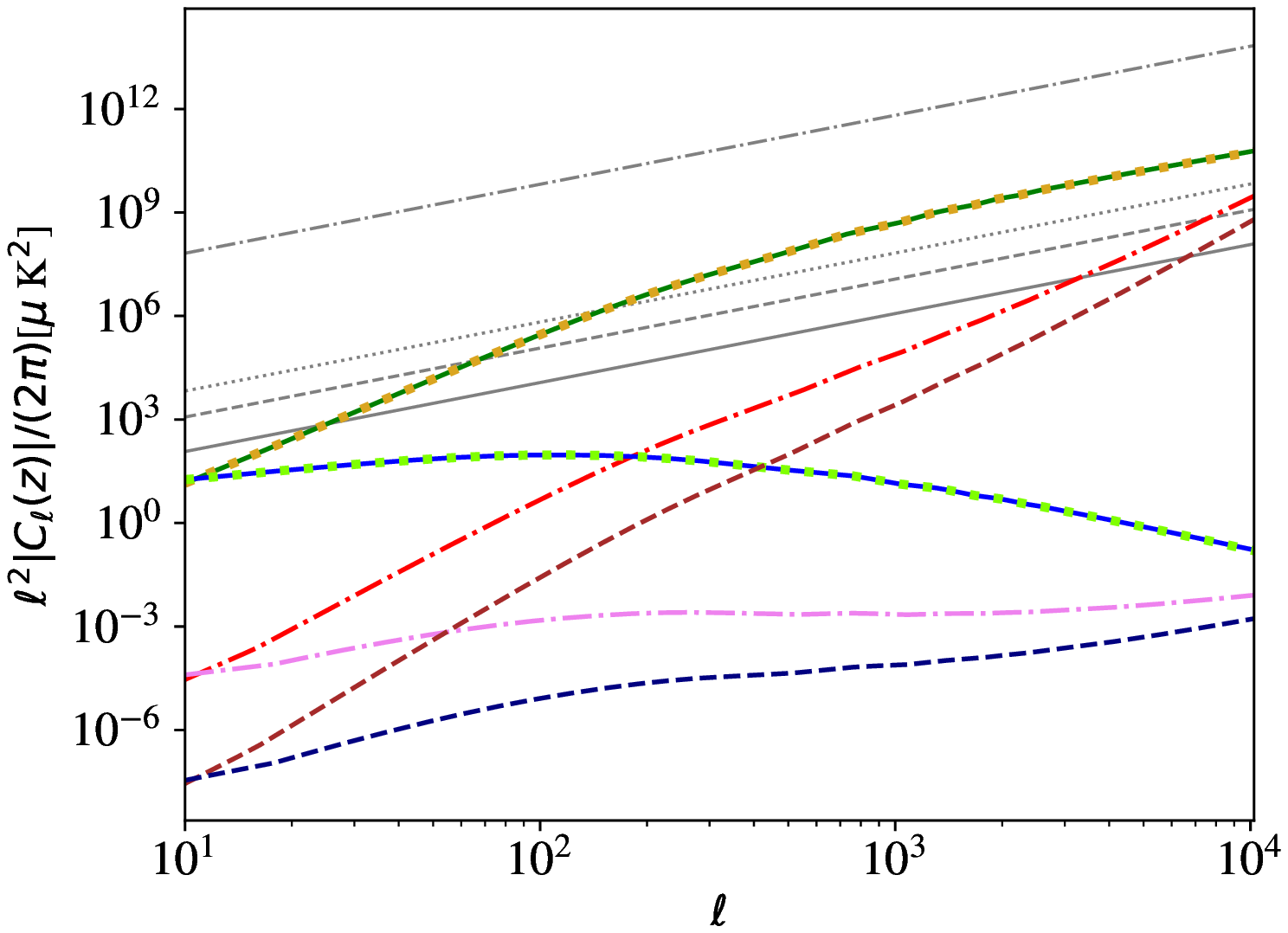}
\epsfxsize=1.9in\epsfbox{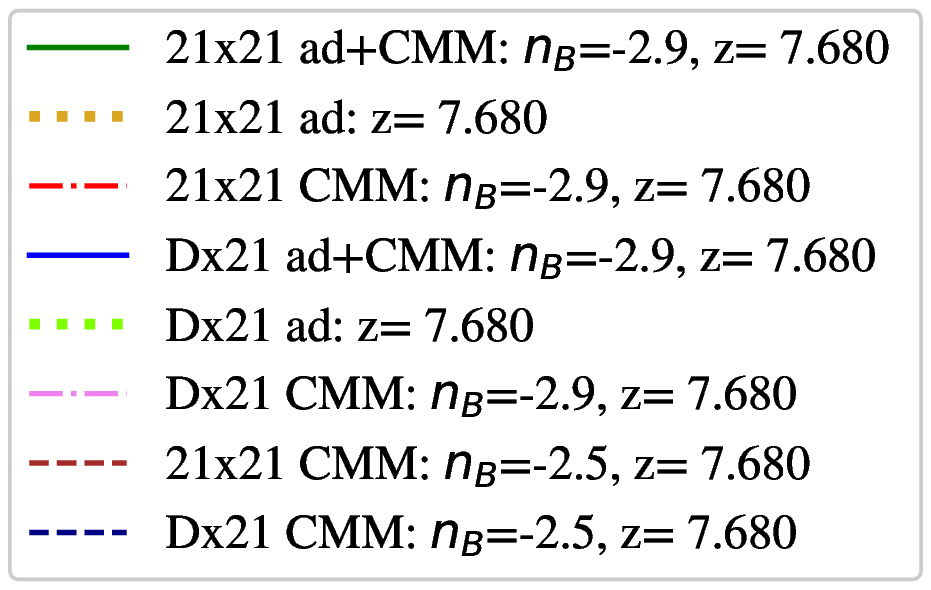}
\epsfxsize=2.8in\epsfbox{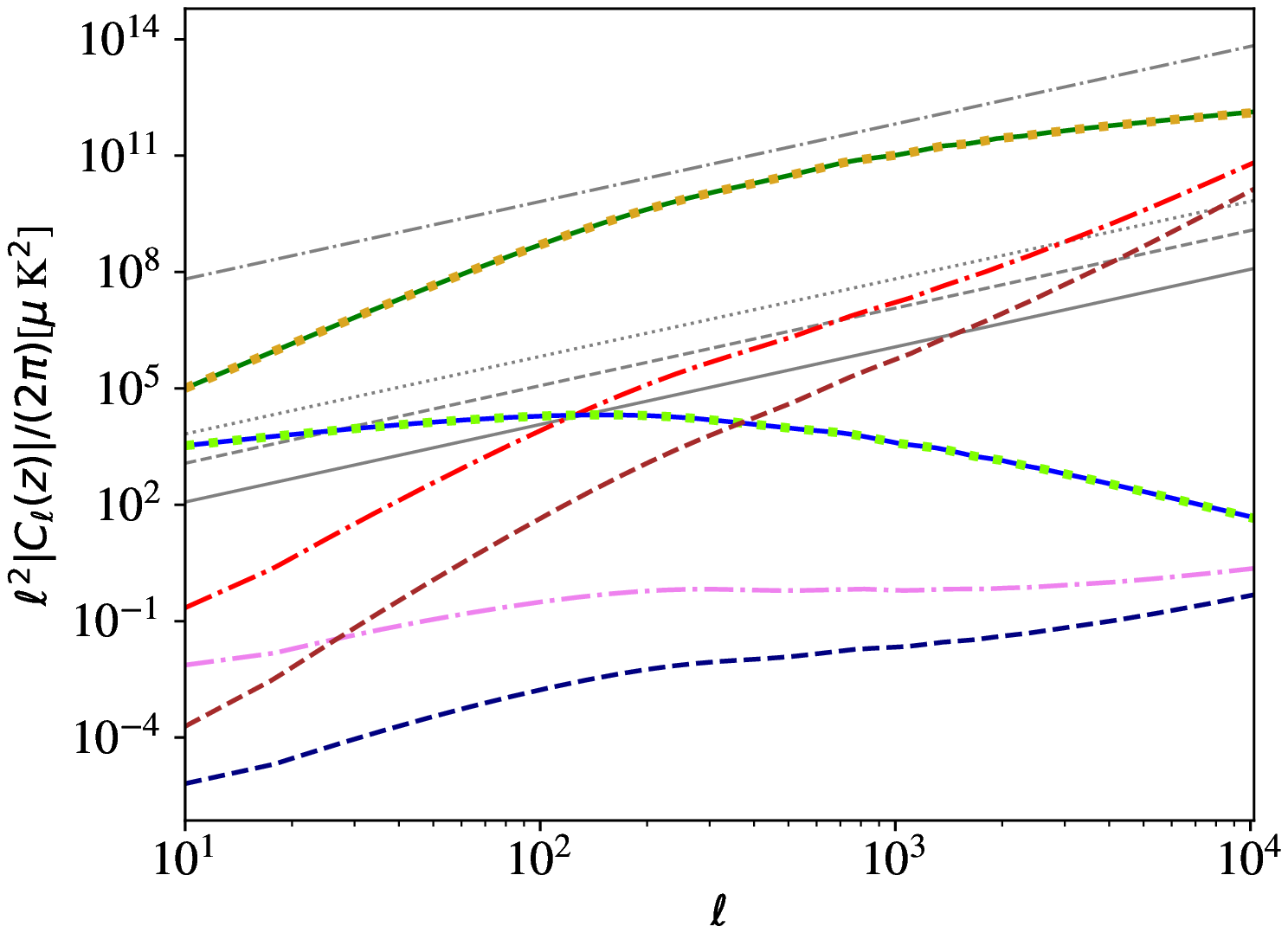}
}
\caption{Absolute values of the angular power spectrum of the auto correlation of the 21 cm line signal (21x21) as well as its cross correlation with 
 the CMB Doppler mode (Dx21)  as a function of multipole $\ell$ for homogeneous ({\sl left}) and inhomogeneous reionization ({\sl right}) for magnetic fields with amplitude $B_0=4$ nG and magnetic spectral indices $n_B=-2.9$ and $n_B=-2.5$ at a reionization redshift $z_{reio}=7.68$. Numerical solutions are shown for the adiabatic mode (ad) and the compensated magnetic mode (CMM) as well as the total (ad+CMM). Also shown are the noise power spectra for LOFAR 
( {\sl grey, dot-dashed line}), SKA1-low  ({\sl grey, dotted line}), SKA1-mid  ({\sl grey, dashed line}) and the optimized configuration 
 SKA1-mid opt ({\sl grey, solid line})  . }
\label{fig1}
\end{figure}

In figure \ref{fig1} the angular power spectra for the auto correlation of the 21 cm line signal as well as its cross correlation for the primary CMB Doppler mode are shown for homogeneous ({\sl left-hand side panel}) and inhomogeneous ({\sl right-hand side panel})  reionization for the adiabatic mode (ad), the compensated magnetic mode (CMM) and the total contribution for magnetic fields with strength $B_0=4$ nG and spectral indices $n_B=-2.9$ and $n_B=-2.5$, respectively.
Moreover,  using  the general expression of 
\cite{Zaldarriaga:2003du, Alvarez:2005sa} for the noise power spectrum,
\begin{eqnarray}
\frac{\ell^2  N_{\ell}^{21}}{2\pi}=\frac{(130 \mu K)^2}{N_{month}\Delta\nu_{\rm MHz}}
\left[
\left(\frac{\ell}{100}\right)
\left(\frac{1+z}{10}\right)\left(\frac{D}{1{\rm km}}\right)\left(\frac{10^3 {\rm m}^2{\rm K}^{-1}}
{A_{eff}/T_{sys}}\right)
\right]^2
\end{eqnarray}
the sensitivities for an observation time of 1.3 months  with LOFAR(4,100,61), SKA1-low(300,80,559) and SKA1-mid(770,150,1560) \cite{SKAtelecon} are also included in figure \ref{fig1}. In parentheses are given the specifications of the, already existing (LOFAR) or in the stage of construction (SKAO),  radio telescope arrays, namely, the bandwidth in MHz, the baseline in km and  the ratio of the effective area over system temperature $A_{eff}/T_{sys}$ in m$^2$/K. 
For SKA1-mid numerical solutions are also shown for 13 months of observation time  as a hypothetical  optimized configuration denoted as SKA1-mid opt.

\begin{figure}[h!]
\centerline{\epsfxsize=2.8in\epsfbox{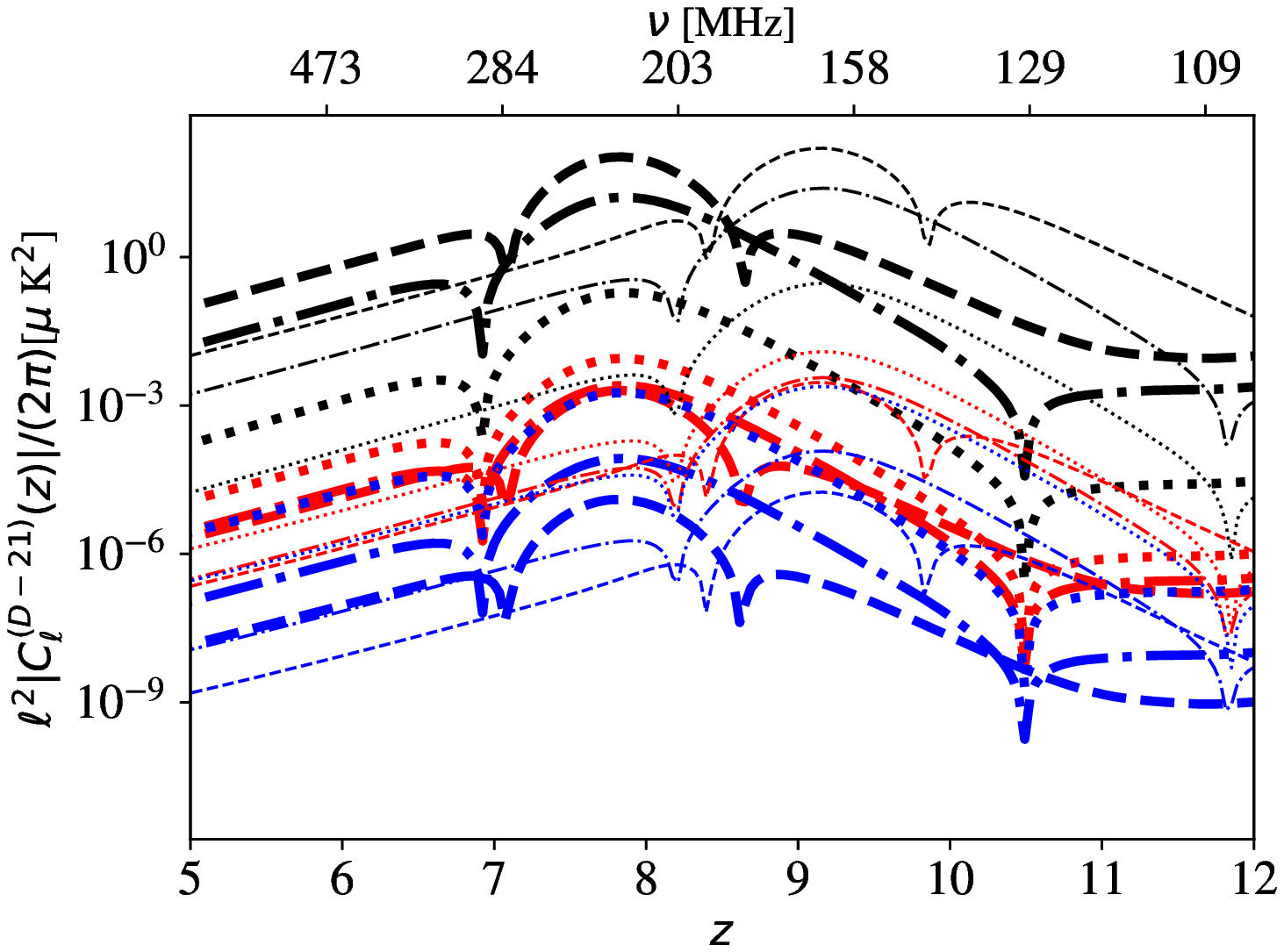}
\epsfxsize=1.9in\epsfbox{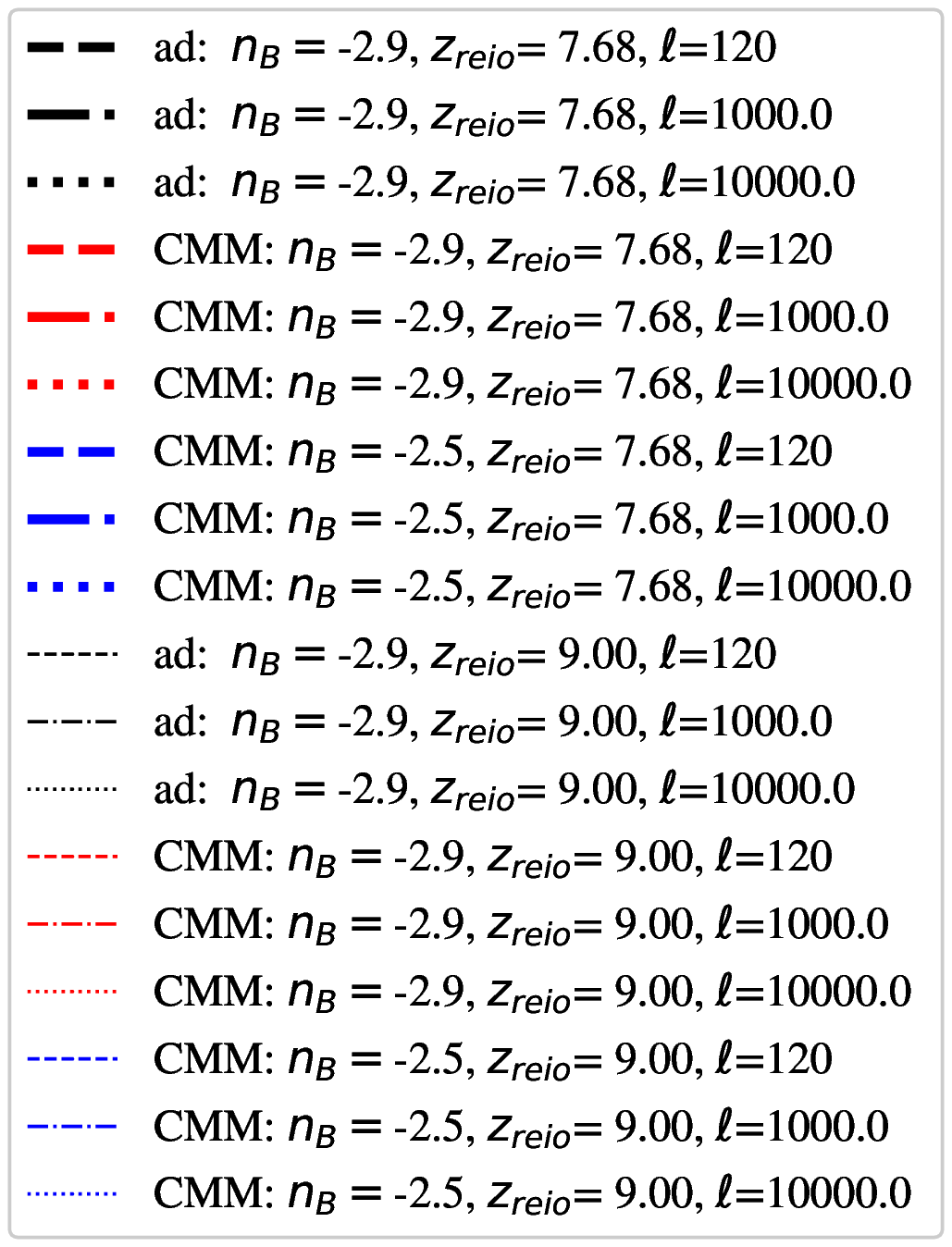}
\epsfxsize=2.8in\epsfbox{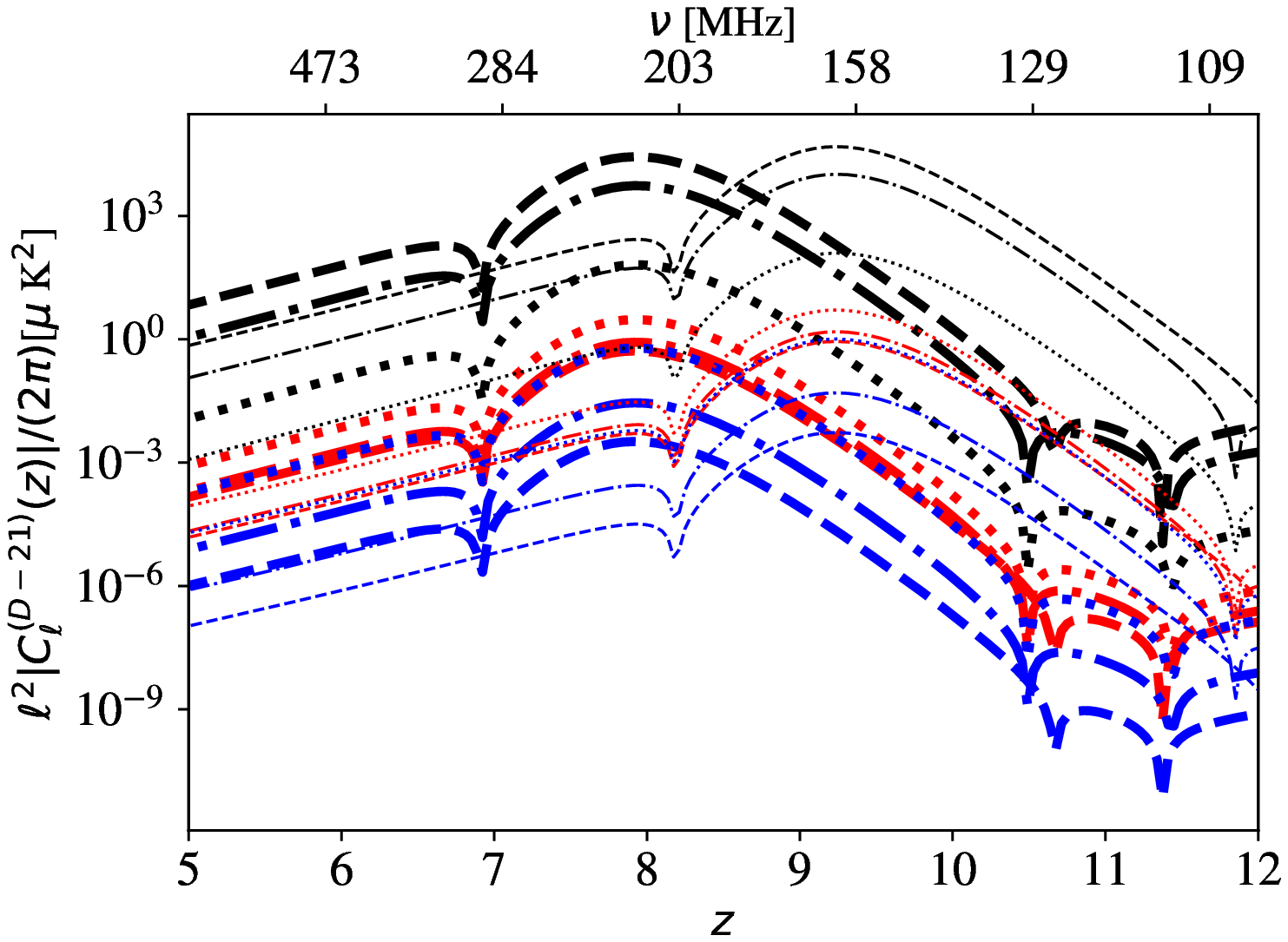}}
\caption{The absolute value of the cross correlation function of the CMB Doppler mode and 21 cm line signal  at multipole $\ell=120, 10^3$ and $\ell=10^4$  as a function of redshift for homogeneous ({\sl left}) and inhomogeneous reionization ({\sl right}) for magnetic fields with amplitude $B_0=4$ nG and magnetic spectral indices $n_B=-2.9$ and $n_B=-2.5$ at reionization redshifts $z_{reio}=7.68$ ({\sl thick lines}) and $z_{reio}=9$ ({\sl thin lines}). Numerical solutions are shown for the adiabatic mode (ad) and the compensated magnetic mode (CMM) .}
\label{fig2}
\end{figure}

The cross correlation function at multipoles $\ell=120, 10^3$ and $10^4$ is shown in figure \ref{fig2} as a function of redshift for two different reionization redshifts, namely the bestfit value of the Planck 2018 only data, $z_{reio}=7.68$ ({\sl thick lines}) and at a higher redshift of $z_{reio}=9$ ({\sl thin lines}). As can be seen in figure \ref{fig2} there is a local extremum around the redshift of reionization. 
The maximal contribution of the cross correlation is found for the adiabatic mode (ad) at the lowest multipole shown, namely, $\ell=120$ which corresponds approximately to the peak in the Doppler mode auto correlation function. This can also be seen in figure \ref{fig1} where the cross correlation function as a function of multipole $\ell$ is shown at $z_{reio}=7.68$.
For the compensated magnetic mode (CMM) the largest contribution of the cross correlation function is found at the largest value,  $\ell=10^4$,  which is due to a local maximum in the linear matter power spectrum generated by the Lorentz term contribution in the baryon velocity evolution equation (e.g., \cite{sl,kk2019, kk21}). Moreover, the largest contribution is found for smaller magnetic field spectral index $n_B=-2.9$ which is reflected in the complementary figure \ref{fig1} showing the evolution of the cross correlation function as a function of $\ell$ at fixed redshift $z=z_{reio}=7.68$.
Comparing the results for homogeneous and inhomogeneous reionization amplitudes are larger for the cross- and auto correlation functions for the latter (cf. figures \ref{fig1} and \ref{fig2}, left- and right-hand side panels, respectively). This is due to the assumption that the fluctuations in the ionization fraction for the inhomogeneous scenario is directly proportional to the matter power spectrum (cf. equation (\ref{deltax}))  leading to additional contributions in the auto- and cross correlation angular power spectra, cf. equations (\ref{C21-21}) and (\ref{CD-21}), respectively.
In figure \ref{fig2} the upper horizontal axis shows the frequency of observation of a 21 cm signal corresponding to 1420 MHz originating at redshift $z$,
namely, $\nu=1420$  MHz$/(1+z)$.

\begin{figure}[h!]
\centerline{\epsfxsize=2.8in\epsfbox{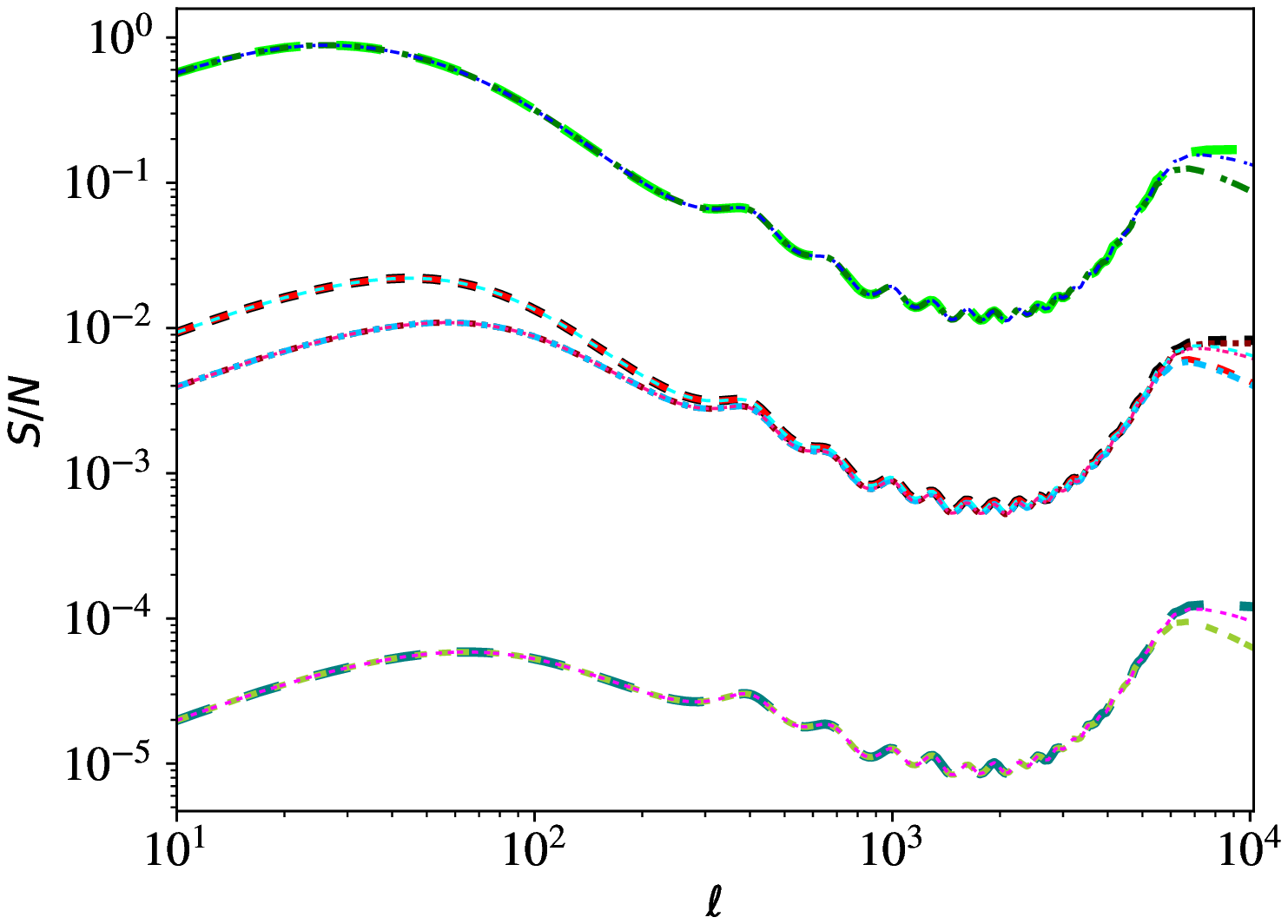}
\epsfxsize=1.9in\epsfbox{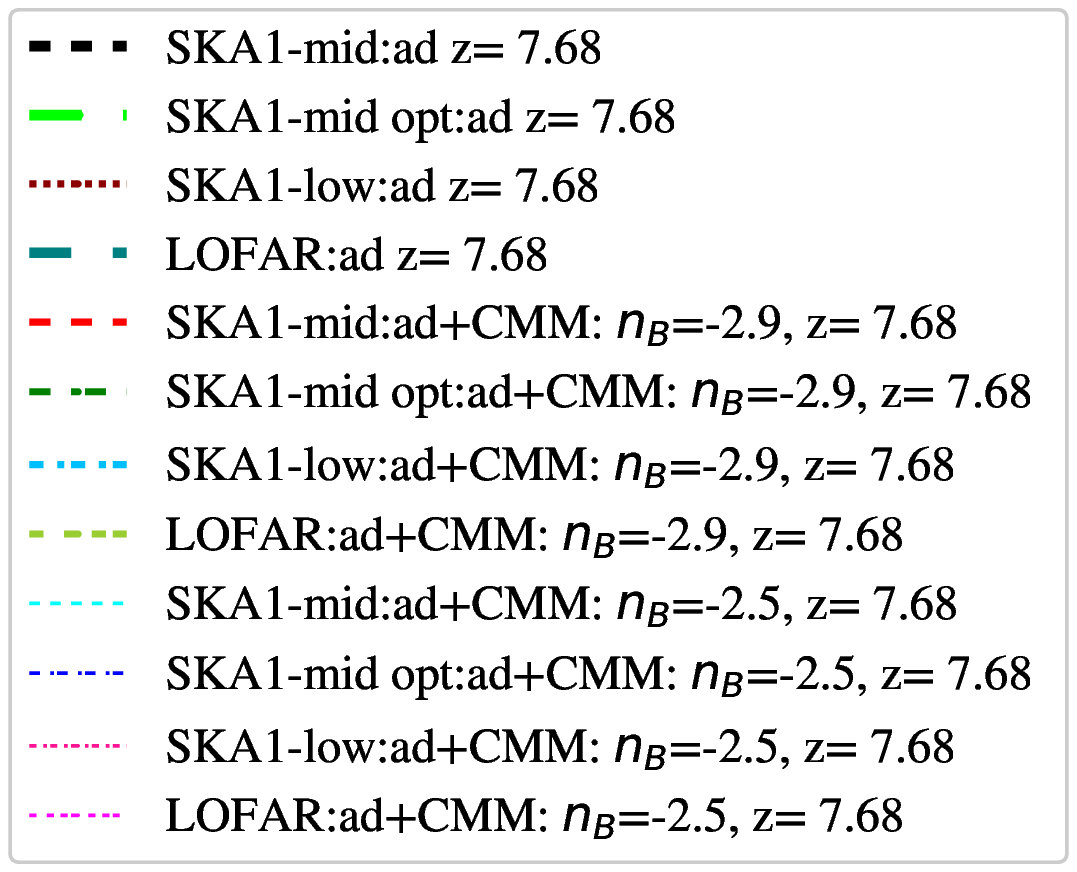}
\epsfxsize=2.8in\epsfbox{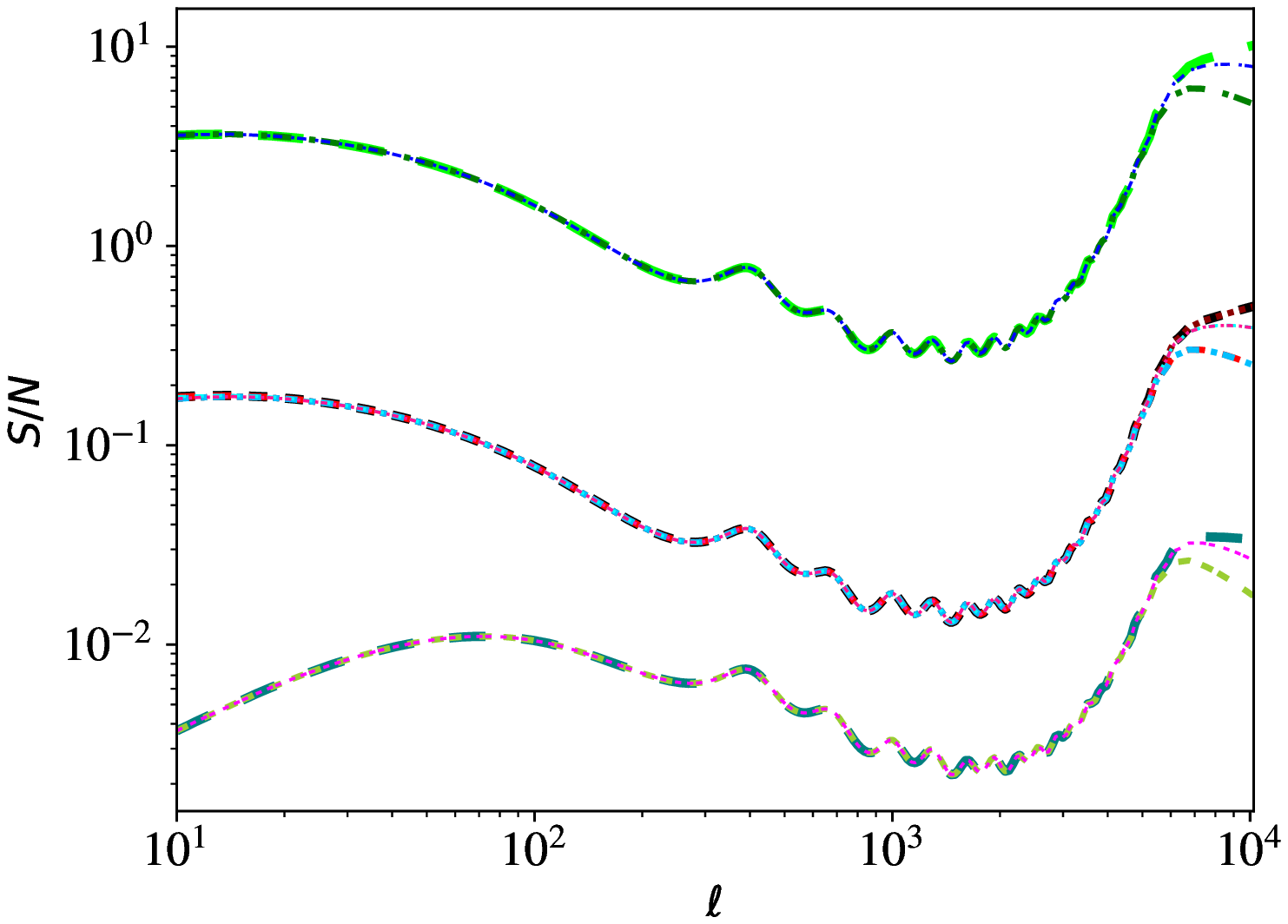}}
\caption{Signal-over-noise ratio of the angular power spectrum of the cross correlation between the CMB Doppler mode and the 21 cm line signal as a function of multipole $\ell$ for homogeneous ({\sl left}) and inhomogeneous reionization ({\sl right}) for magnetic fields with amplitude $B_0=4$ nG and magnetic spectral indices $n_B=-2.9$ and $n_B=-2.5$ at reionization redshift $z_{reio}=7.68$. Numerical solutions are shown for the adiabatic mode (ad) and the compensated magnetic mode (CMM) as well as the total (ad+CMM) for LOFAR, SKA1-low, SKA1-mid and SKA1-mid opt.}
\label{fig4}
\end{figure}

For the potential observability the signal-over-noise ratio $\frac{S}{N}$ is an important indicator. 
This can be estimated as \cite{Ma,Dore,Adshead:2007ij}
\begin{eqnarray}
\left(\frac{S}{N}\right)^2=\frac{f_{sky}(2\ell+1)\Delta\ell_{bin}\left(C_{\ell}^{(D-21)}\right)^2}
{C_{\ell}^{(CMB)}\left(C_{\ell}^{(21-21)}+N_{\ell}^{(21)}\right)
+\left(C_{\ell}^{(D-21)}\right)^2},
\end{eqnarray}
where $\Delta\ell_{bin}\simeq 0.46\ell$ is the bin width at a given $\ell$ and $f_{sky}$ the observed sky fraction observed by both the CMB and 21 cm line telescope arrays. Following \cite{Ma} $f_{sky}$ is set to $f_{sky}^{CMB,LOFAR}=0.0006$ for CMB and LOFAR observations and $f_{sky}^{CMB,SKA}=0.0024$  for CMB and SKA observations. Also included is the hypothetical optimized configuration and survey design for 13 months of observation and $f_{sky}^{CMB,opt}=1$ denoted as SKA1-mid opt.
In figure \ref{fig4} the signal-over-noise ratio is shown for LOFAR, SKA1-low, SKA1-mid as well as SKA1-mid opt.
The signal-over-noise ratio only becomes of ${\cal O}(1)$ for the hypothetical optimized configuration SKA1-mid opt
in the case of homogeneous reionization (cf. figure \ref{fig4}, left-hand panel) and $\frac{S}{N}>1$ for inhomogeneous reionization (cf. figure \ref{fig4}, right-hand panel). However, in the latter case it is interesting to note that at high multipoles solutions for different magnetic field parameters are distinguishable. Moreover, even the results for the proposed SKA1-mid configuration are in general not much smaller than one implying that an adjustment of the planned survey and configuration specifications might lead to $\frac{S}{N}\sim 1$. Thus this might open up the possibility to constrain the magnetic field parameters.

\begin{figure}[h!]
\centerline{\epsfxsize=2.8in\epsfbox{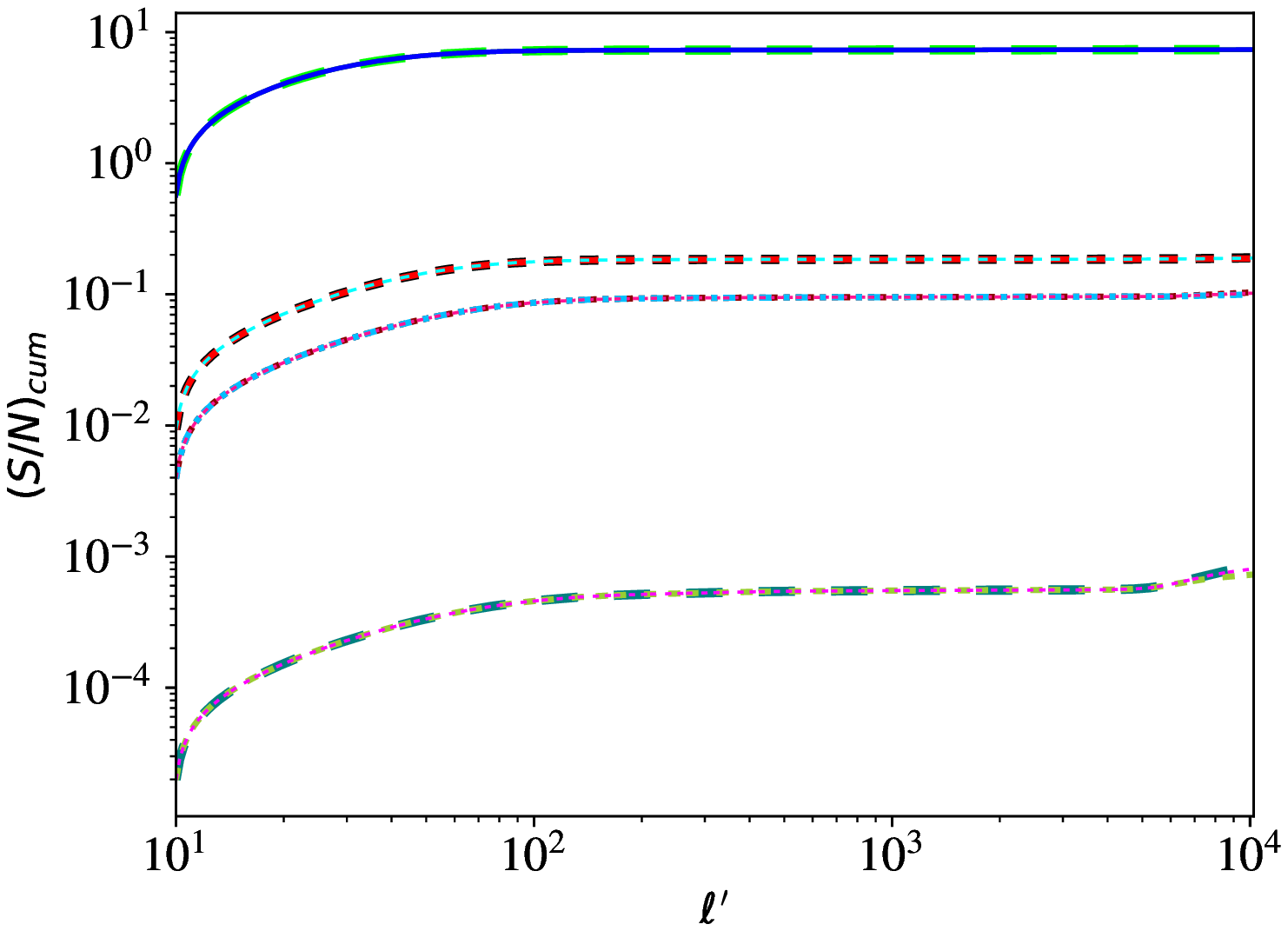}
\epsfxsize=1.9in\epsfbox{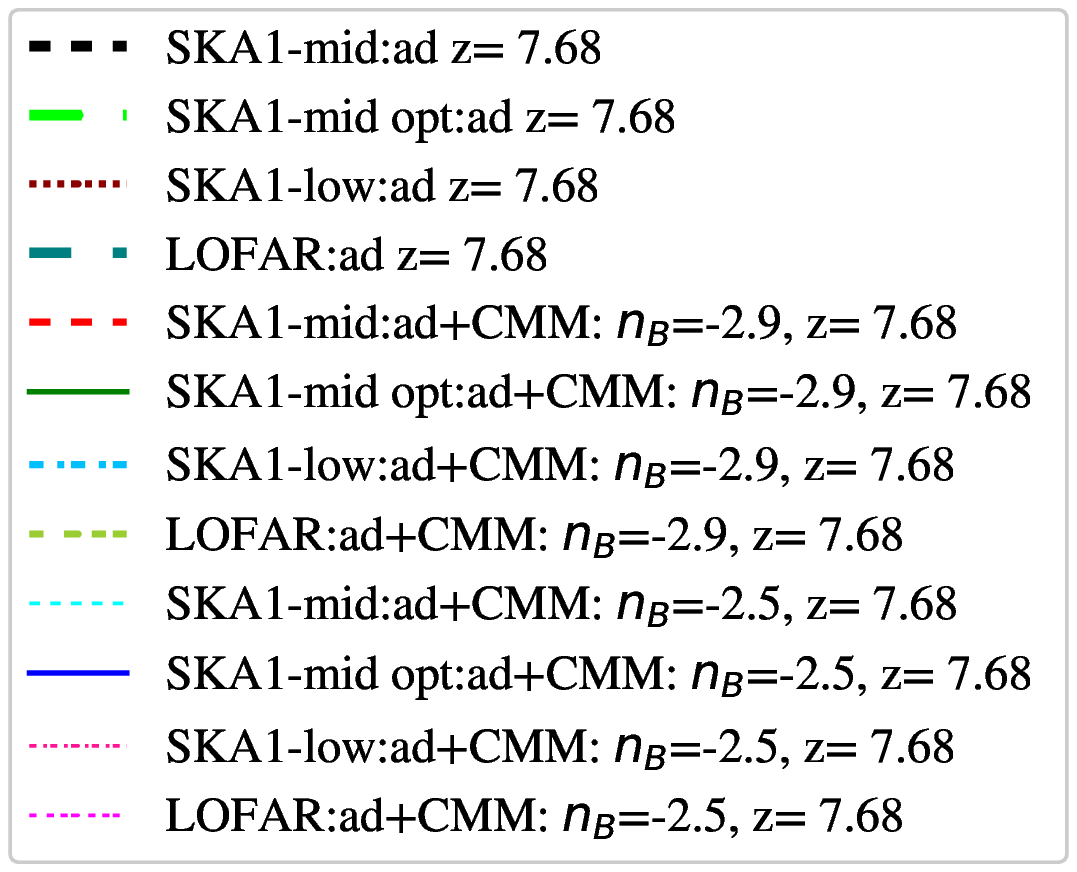}
\epsfxsize=2.8in\epsfbox{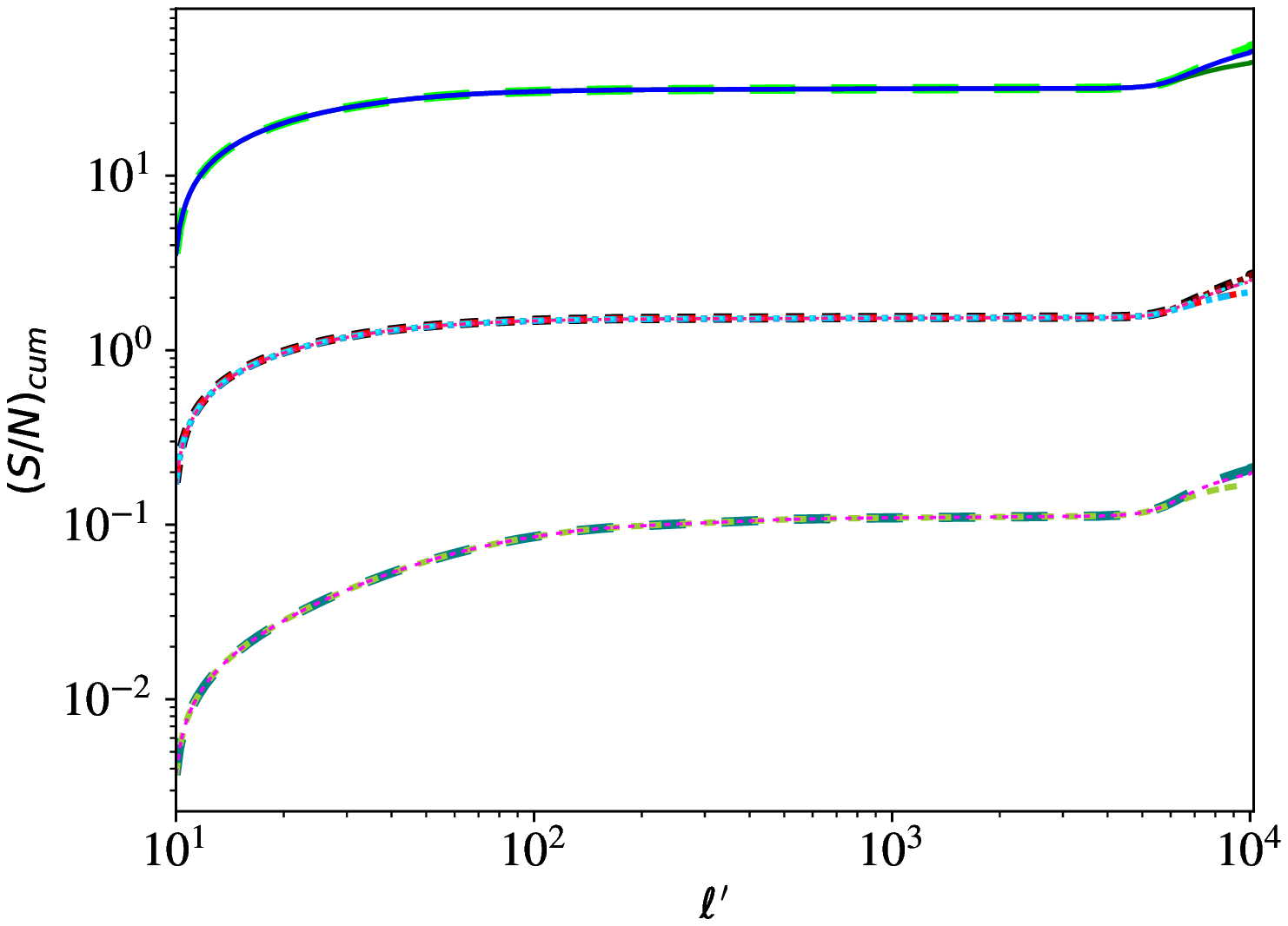}}
\caption{Cumulative signal-over-noise ratio of the angular power spectrum of the cross correlation between the CMB Doppler mode and the 21 cm line signal as a function of the maximal multipole $\ell'$ for homogeneous ({\sl left}) and inhomogeneous reionization ({\sl right}) for magnetic fields with amplitude $B_0=4$ nG and magnetic spectral indices $n_B=-2.9$ and $n_B=-2.5$ at reionization redshift $z_{reio}=7.68$. Numerical solutions are shown for the adiabatic mode (ad) and the compensated magnetic mode (CMM) as well as the total (ad+CMM) for LOFAR, SKA1-low, SKA1-mid and SKA1-mid opt.}
\label{fig5}
\end{figure}

It is also interesting to consider the cumulative signal-over-noise ratio defined by $\left(\frac{S}{N}\right)_{cum}$ \cite{Ma}
\begin{eqnarray}
\left(\frac{S}{N}\right)_{cum}^2(<\ell')=\sum_i
\frac{f_{sky}(2\ell_i+1)\Delta\ell_{bin,i}\left(C_{\ell_i}^{(D-21)}\right)^2}
{C_{\ell_i}^{(CMB)}\left(C_{\ell_i}^{(21-21)}+N_{\ell_i}^{(21)}\right)
+\left(C_{\ell_i}^{(D-21)}\right)^2},
\end{eqnarray}
where $i$ denotes the $ith$ bin with  $\ell_i$ its central multipole and $\ell'$ the maximal multipole implying $\ell_i\leq \ell'$.
In figure \ref{fig5} the cumulative signal-over-noise ratio as a function of the maximal multipole $\ell'$
is shown for  LOFAR, SKA1-low, SKA1-mid as well as SKA1-mid opt.
For  homogeneous reionization the cumulative signal-over-noise ratio is less than one in all but the case of the SKA1-mid opt configuration where it is of ${\cal O}(10)$ (cf. figure \ref{fig5} ({\sl left-hand side panel})). As can be seen in figure \ref{fig5} ({\sl right-hand side  panel}) for inhomogeneous reionization  for $\ell'>20$ configurations of SKA1-low and  SKA1-mid  have a cumulative signal-over-noise ratio of order one and larger  and the SKA1-mid opt configuration  has values between 30 and 50 for 
$\left(\frac{S}{N}\right)_{cum}$.  Moreover, there is a distinction between solutions for different magnetic field parameters for large values of the maximal multipole $\ell'$.

\begin{figure}[h!]
\centerline{\epsfxsize=2.8in\epsfbox{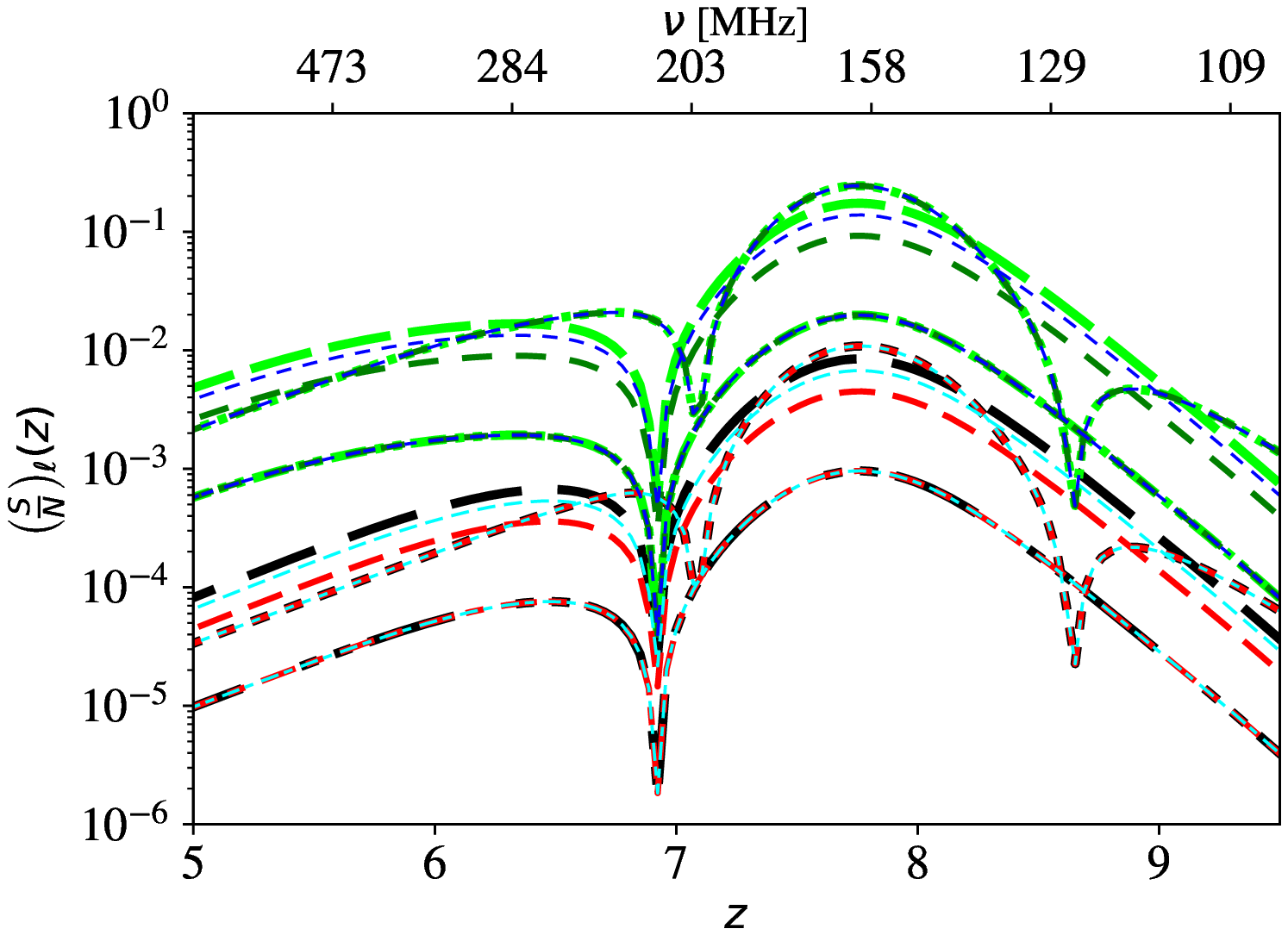}
\epsfxsize=1.9in\epsfbox{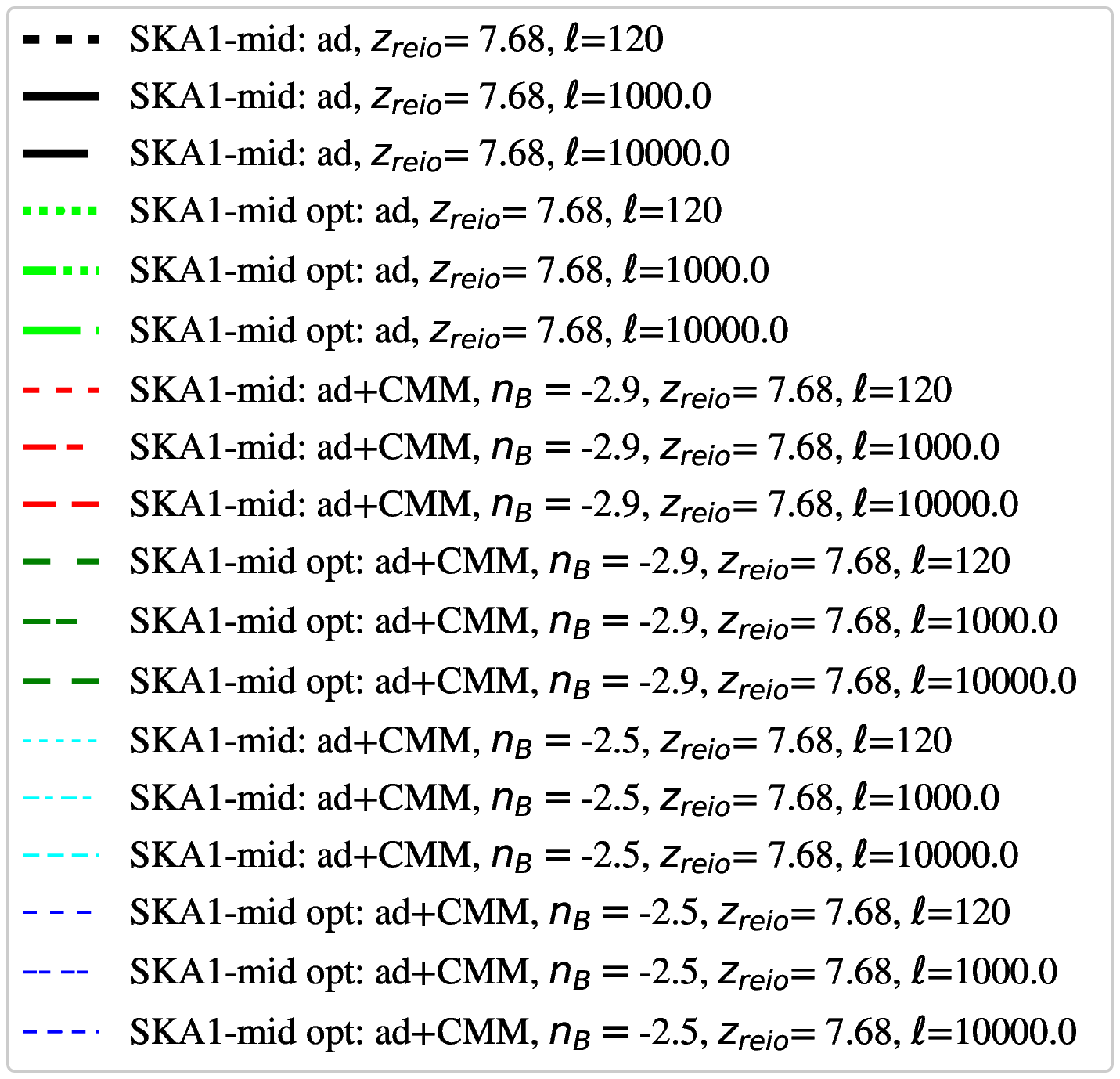}
\epsfxsize=2.8in\epsfbox{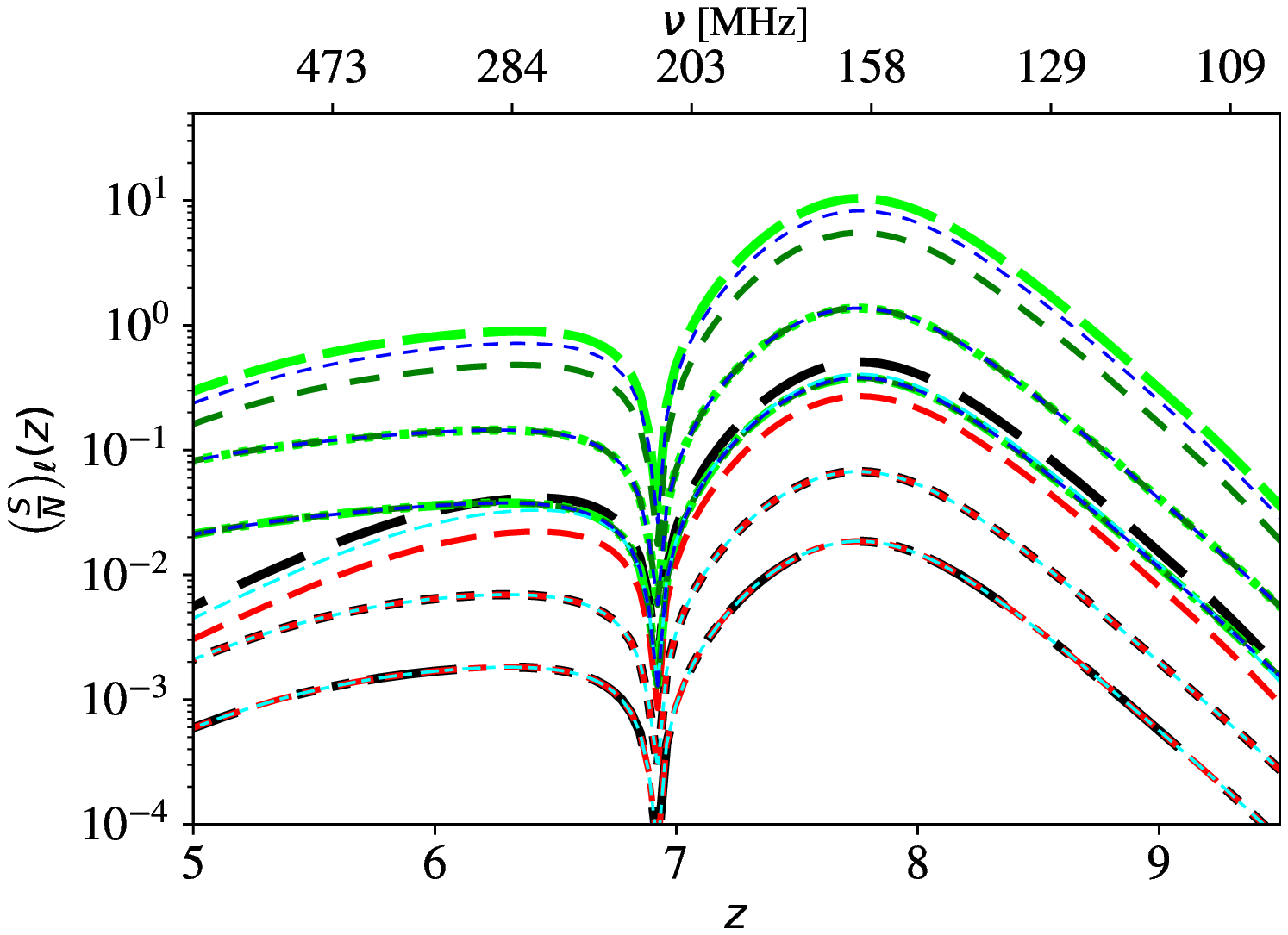}}
\caption{Signal-over-noise ratio of the angular power spectrum of the cross correlation between the CMB Doppler mode and the 21 cm line signal as a function of  redshift $z$ for multipoles $\ell=120,10^3, 10^4$  for homogeneous ({\sl left}) and inhomogeneous reionization ({\sl right}) for magnetic fields with amplitude $B_0=4$ nG and magnetic spectral indices $n_B=-2.9$ and $n_B=-2.5$ at reionization redshift $z_{reio}=7.68$. Numerical solutions are shown for the adiabatic mode (ad) and the compensated magnetic mode (CMM) as well as the total (ad+CMM) for  SKA1-mid and SKA1-mid opt.}
\label{fig6}
\end{figure}

In figure \ref{fig6} the signal-over-noise ratio $\left(\frac{S}{N}\right)_{\ell}(z)$ is shown  as a function of redshift at fixed different multipoles $\ell=120, 10^3, 10^4$.
The reionization redshift is set to $z_{reio}=7.68$ corresponding to the bestfit value of the Planck 2018 data.
$\left(\frac{S}{N}\right)_{\ell}(z)$ 
is only shown for SKA1-mid and the
hypothetical  optimized configuration SKA1-mid opt. This is motivated by the previous results.
As can be seen in  figure \ref{fig1} the amplitude of the noise angular power spectrum is smallest for the SKA1-mid configurations. In figures \ref{fig4}
and \ref{fig5} the signal-to-noise ratio as function of $\ell$ at a fixed redshift as well as its cumulative value are largest  for the SKA1-mid configurations indicating the best prospects for detection.
As can be seen in figure \ref{fig6} the signal-over-noise ratios indicate a distinction between the pure adiabatic mode and the total adiabatic plus compensated magnetic mode for magnetic field parameters $B_0=4$ nG and $n_B=-2.9$ and $n_B=-2.5$, respectively. In particular, considering the case of inhomogeneous reionization amplifies this separation of curves. Therefore, considering different survey designs might be interesting in order to use the cross correlation between the CMB Doppler mode temperature fluctuations and the 21 cm line brightness fluctuations to constrain parameters of a probable cosmological magnetic field.

\section{Conclusions}
\label{sec3}
\setcounter{equation}{0}

The cross correlation angular power spectrum between the CMB Doppler mode and the 21 cm line signal has been calculated in the presence of a primordial magnetic field assumed to be a random gaussian non helical field. Numerical solutions have been obtained for homogeneous and inhomogeneous reionization for CMB observations and 21 cm line radio telescope configurations and surveys with LOFAR and SKA1-low as well as SKA1-mid. In the signal-over-noise $S/N$ figures also a hypothetical  optimized configuration of SKA1-mid has been included. Solutions for $S/N$ have been presented as a function of multipole $\ell$ at fixed redshifts $z$ as well as a function of redshift $z$ at fixed values of multipole $\ell$.
For magnetic fields with amplitude $B_0=4$ nG and spectral indices $n_B=-2.9$ and $n_B=-2.5$, respectively, the signal-over-noise ratios resulting from the adiabatic and magnetic mode contributions for inhomogeneous reionization reaches values larger than  1 and in the case of the cumulative signal-over-noise ratio 
it reaches values between 30 and 50. However, it should be taken into account that these results depend on the particular model of reionization used here.  
These figures indicate that there might be potentially a possibility for advanced configurations to constrain magnetic field parameters with 21 cm line observations.


\section{Acknowledgements}

Financial support by Spanish Science Ministry grant PID2021-123703NB-C22 (MCIU/AEI/FEDER, EU) and Basque Government grant IT1628-22 is gratefully acknowledged.



\bibliography{references}

\bibliographystyle{apsrev}

\end{document}